\documentclass[twocolumn,secnumarabic,amssymb, nobibnotes, aps, prd,nofootinbib]{revtex4-2}

\usepackage[...]{youngtab}
\usepackage{xfrac}
\usepackage{appendix}
\usepackage{amsfonts}
\usepackage{amssymb}
\usepackage{physics}
\usepackage[usenames,dvipsnames]{xcolor}
\usepackage{tabularx,multirow,array,amsmath,mathalfa}
\usepackage{color}
\usepackage{ccaption}
\usepackage{mathtools}
\usepackage{bbold}
\usepackage{kantlipsum}
\usepackage{tikz}
\usepackage[graphicx]{realboxes}

\usepackage{pgfplots}
\newcommand{\mdot}{\raise1.5pt \hbox{.}}
\usepackage{braids}
\newcommand{\be}{\begin{equation}}
\newcommand{\ee}{\end{equation}}

\begin{document}

\title{Knot-Quiver correspondence for double twist knots}

\author{Vivek Kumar Singh}%
 \email{vks2024@nyu.edu}
\affiliation{Center for Quantum and Topological Systems (CQTS), NYUAD Research Institute, New York University Abu Dhabi, PO Box 129188, Abu Dhabi, UAE  
}%

\author{Sachin Chauhan}%
\affiliation{Department of Physics, Indian Institute of Technology Bombay, Powai, Mumbai, 400076, India  
}%

\author{Aditya Dwivedi}
\affiliation{%
Department of Physics, Institute of Science, Banaras Hindu University, Varanasi, 221005, India 
}%
\author{P. Ramadevi}
\affiliation{%
Department of Physics, Indian Institute of Technology Bombay, Powai, Mumbai, 400076, India 
}%
\author{B. P. Mandal}
\affiliation{%
Department of Physics, Institute of Science, Banaras Hindu University, Varanasi, 221005, India 
}%
\author{Siddharth Dwivedi}
\affiliation{Department of Physics, School of Physical Sciences, Central University of Rajasthan, Ajmer, India 
}%

\begin{abstract}
We obtain  a quiver representation for a family of knots called double twist knots $K(p,-m)$. Mainly, we exploit  the reverse engineering of  Melvin-Morton-Rozansky(MMR) formalism to deduce the pattern of the matrix for these quivers.

\end{abstract}

\maketitle
\tableofcontents

\section{Introduction}

Knot-quiver correspondence (KQC) conjectured by Kucharski-Reineke-Stosic-Sulkowski\cite{Kucharski:2017ogk} provides a new encoding of HOMFLY-PT invariants of knots  in terms of the representation theory of quivers. Such a correspondence was motivated by studying the supersymmetric quiver quantum mechanics description of BPS states in brane systems describing knots\cite{PhysRevD.96.121902}.
 
  Quivers are denoted as  directed graphs with finite number of vertices connected by oriented edges. For a quiver with $n$ number of vertices, the directed graph is encoded in a $n\times n$ quiver matrix $C$.  The diagonal elements $C_{ii}$  refer to the number of loops at the `$i$'-th vertex, and the off-diagonal elements $C_{ij}$  give the  number of oriented edges from the vertex `$i$' to the vertex `$j$'. Hence, the elements in the matrix $C$ are non-negative integers.
  
 According to the conjecture, at least one  quiver graph $Q_K$ is associated with the knot $K$ satisfying the exponential growth property of
 the symmetric $r$-colored HOMFLY-PT polynomials $\bar{P}_r^K(A,q)$ (normalized appropriately)\footnote{This normalization includes dividing the reduced $r$-colored HOMFLY-PT polynomial ($P_r^K(A,q)$ in variables $A$ and $q$) by 
 $q$-Pochhammer $(q^2; q^2)_r$ (also known as intermediate normalization).} as elaborated in the Refs.\cite{Kucharski:2017ogk,Ekholm:2021gyu,Ekholm:2021irc}. 
 In the context of KQC, the quivers  are {\it symmetric} quivers. That is., $C_{ij}^{(K)}= C_{ji}^{(K)}$. Further, these quiver matrix elements can be negative integers and be made non-negative by change of  {\it framing}.
Particularly, for the knot $K$ which obey exponential growth property, we can write the  generating series for $\bar{P}_r^K(A,q)$ in two equivalent forms:
 \begin{eqnarray} {\label{KQ1}}
 P_{Q_K} (x) &=& \sum_r \bar{P}_r^K(A,q) x^r \nonumber\\
 ~&=&\sum_{\mathbf{d}}(-q)^{\sum_{i,j}^n d_i C_{ij}^{(K)}d_j} \mathbf{x}^{\mathbf{d}}\prod_{i=1}^n\prod_{j=1}^{d_i}\frac{1}{1-q^{2j}}\nonumber\\
 ~&=&\prod_{\mathbf{d}\neq 0}\prod_{j\in \mathbb{Z}}\prod_{k\geq 0}\left(1-(-1)^j\mathbf{x^d}q^{j+2k+1}\right)^{-\Omega_{\mathbf{d},j}},\nonumber\\
 \end{eqnarray}
 to extract the quiver matrix $C^{(K)}$ as well as the motivic Donaldson-Thomas (DT) invariants
 $\Omega_{\mathbf{d},j}$. Here ${\mathbf{d}}\equiv (d_1,d_2, \ldots d_n)\geq 0$ and $\mathbf{x^d}= \prod_{i=1}^n x_i^{d_i}$ with $x_i=x A^{\beta_i} q^{\alpha_i-1}(-1)^{\gamma_i}$. Note that sets $\{d_i\}$ satisfy the condition $r = d_1 + d_2 + \ldots + d_n$. The procedure to obtain such a motivic series for any knot is still an open question.
 
  For a class of torus knots $(2, 2p+1)$, the twist knots $K_p$($p \in \mathbb Z$) and other knots  upto seven crossings the quiver presentation were obtained\cite{Kucharski:2017ogk,Stosic:2017wno}. Except for unknot and trefoil, knots have more than one quiver presentation with the same number of nodes, indicating that the correspondence of knots to quivers is not unique. In Ref. \cite{PhysRevD.104.086017},  equivalent  quivers with the same number of nodes were shown as vertices on a permutohedra graph, giving a systematic enumeration of such equivalent quivers. There are also quivers with different number of nodes that describe the same physics, i.e. a pool of dualities in 3d $\mathcal{N}=2$ theory\cite{Ekholm:2019lmb}.

The  physical and the geometrical interpretation of the conjectural KQC 
 was addressed \cite{Ekholm:2018eee} within the framework of  Ooguri-Vafa large $N$ duality. From the physics perspective, the motivic generating series of $Q_K$  matches the vortex partition function of 3d $\mathcal{N}=2$ theory $T[Q_K]$.  On the geometrical side\cite{Ekholm:2018eee}, the spectrum of holomorphic curves with boundary on the conormal Lagrangian $L_K$ of the knot in the resolved conifold encodes the quiver data. That is., the basic holomorphic disks correspond to the nodes of the quiver $Q_K$ and the linking of their boundaries to the quiver arrows. 

With double fat diagram description for arborescent knots, the $r$-colored HOMFLY-PT invariant: $P_r^{\mathcal{K}}(A=q^N,q)$ ($r$ refers to symmetric color $\underbrace{\yng(4)}_r \in SU(N)$)  for a universal arborescent knot family $\mathcal K$ involving seven parameters was proposed in\cite{Mironov:2015aia,Mironov:2016deg}. Such a family includes most of the arborescent knots upto 10 crossings. Also, the colored HOMFLY-PT polynomials (upto four symmetric colors for knots upto 10 crossings) have been updated in the website\cite{knotebook}. Recent papers on the  existence  of quivers for all rational knots, tangles and arborescent knots\cite{Stosic:2020xwn,Stosic:2017wno}  motivated us to deduce a universal quiver  for our arborescent knot family. Even though the problem is concrete, finding explicit quivers for this universal arborescent family appears to be a hard problem. 

As a first step, we  wanted to investigate some arborescent knots whose Alexander polynomial has a  structure similar to that of the twist knots $K_p$. That is., $\Delta(X) =1-pX$. In fact, there is a systematic reverse engineering approach of the Melvin-Morton-Rozansky (MMR) formalism to obtain the quiver representation for such twist knots\cite{Banerjee_2020}. We observed $\Delta(X)= 1-4X$ for knot $8_3$ is part of the family of double twist knots characterized by two variables, denoted as $K(p,-m)$, illustrated in Figure \ref{DT}. Note $p,m \in \mathbb Z_+$ denote the number of full-twists and the Alexander polynomial is $\Delta(X)= 1- pm X$.  Such a form motivated us to attempt quiver representation for double twist knots.

Even though the $r$-colored HOMFLY-PT for any double twist knot in the cyclotomic form are known\cite{chen2021cyclotomic,Wang:2020uz}, rewriting them in the form of motivic series is still a challenging problem. We tried to determine  the quiver representation of $P_r^{K(p,-m)}(A=q^N,q)$  following the methodology in Ref.\cite{Banerjee_2020}. However, we faced computational difficulty in deducing the $A$ dependence in the quiver representation. Also, we know that the quiver matrix $C^{(K)}$  do not depend on the variable $A=q^N$. As our aim is to conjecture the quiver matrix form for the double twist knots $K(p,-m)$,  we focus on rewriting $r$-colored Jones polynomial ($A=q^{N=2}$) :
 $$J_r(K(p,-m),q) \equiv P_r^{K(p,-m)}(A=q^2,q)~,$$
 as a motivic series. Particularly, we  obtain  quiver  matrix $C_{i,j}^{K(p,-m)}$ associated with the $K(p,-m)$ for $m\leq 3$. We conjecture that the quiver matrix $C_{i,j}^{K(m,-m)}$  is sufficient to recursively generate the quiver matrix for all the double twist knots $K(p\neq m,-m)$.

 We follow the route of reverse engineering of MMR expansion\cite{Banerjee_2020}  to derive  the motivic series form for $P_r^{K(p,-m)}(A=q^N,q)\Big |_{N=2}$.
 We will now briefly review the reverse engineering formalism, which will set the notation and procedure we follow for $K(p,-m)$ in the next section.
 
  \subsection{Reverse Engineering of Melvin-Morton-Rozansky (MMR) expansion}  
  Melvin-Morton-Rozansky(MMR) expansion states that  the symmetric $r$-colored  HOMFLY-PT for knot $\mathcal K$ has the following  semiclassical expansion:
\begin{eqnarray} 
\lim_{\hbar\to 0, r\to\infty}  P_r^{\mathcal{K}}(A,q=e^{\hbar})&\simeq& \frac{1}{\Delta^{\mathcal{K}}(x)^{N-1}} + \nonumber\\&&\sum_{k=1}^\infty \left(\frac{R_k^{\mathcal{K}}(x,N)}{\Delta^{\mathcal{K}}(x)}\right)^{N+2k-1} \hbar^k,\nonumber\\
\label{mmr-intro}
\end{eqnarray}
with the leading term being the Alexander polynomial $\Delta^{\mathcal{K}}(x)$ and the variable  $x$ in terms of color $r$ is $x=q^r={\rm ~const}$. The symbol $R^{\mathcal{K}}_k(x, N)$ represent polynomials in the variable $x$. The reverse approach is to obtain 
$P_r^{\mathcal{K}}(A,q)$ using the Alexander polynomial $\Delta^{\mathcal{K}}(x)$\cite{Banerjee_2020}. This approach also has obstacles to lift the $\hbar \rightarrow 0$ expansion to $q$-dependent $P_r^{\mathcal{K}}(A,q)$  but can be fixed for some situations by comparing with the data of symmetric $r$-colored HOMFLY-PT polynomials known for $r=1,2,3$.
We will briefly highlight the steps involved in the reverse engineering formalism of MMR expansion\cite{Banerjee_2020}: 
\begin{enumerate}
    \item[i.] We rewrite the Alexander polynomial in new variable $X=\frac{(1-x)^2}{x}$. Thus, the Alexander polynomial takes the following form:
    \begin{equation*}
        \Delta(x)=1-\sum_{i=1}^{s}a_i\frac{(1-x)^{2i}}{x^i}\equiv 1-g(X)
    \end{equation*}
where the coefficients $a_i$ are integers, and $s$ is a positive integer.
    \item[ii.] Now, we use the following inverse binomial theorem \begin{equation*}
        \frac{1}{(1-u)^n}=\sum_{m=0}^{\infty}\binom{n+m-1}{m}u^m
    \end{equation*}
to write the first term of MMR expansion (\ref{mmr-intro}) as follows:
\begin{eqnarray}{\label{MMRE}}
    \frac{1}{\Delta(x)^{N-1}}&=&\sum_{m=0}^{\infty}\binom{N+m-2}{m}g(X)^m=\sum_{k=0}^{\infty}c_k X^k\nonumber\\
   &=& \sum_{k=0}^{\infty}(k! c_k)\frac{X^k}{k!}
\end{eqnarray}
\item[iii.] We make the following quantum deformation to get the quantum-deformed polynomial:\\
    \begin{equation*}
        \frac{X^k}{k!}=\frac{(1-x)^{2k}}{k!x^k}\rightsquigarrow {r \brack k}_q q^{-rk}(-A q^rt^3;q)_k.
    \end{equation*}
Here, the variable $t$ is known as the refined parameter. In this article, we will take $t = -1$ to obtain unrefined polynomial invariants for double twist knots.  The term within parentheses represents the $q$-Pochhammer, while square brackets correspond to the q-binomials, which are defined as: 
    $${r \brack k}_q=\frac{(q;q)_r}{(q;q)_k(q;q)_{r-k}}~~~;~~~ (x;q)_k=\prod_{i=0}^{k-1}(1-x q^{i}).$$
\item[iv.] Further, the coefficient $k!c_k\xrightarrow{\rm q-deformation} \Tilde{c}^{\mathcal{K}}_k$ depends on the knot $\mathcal{K}$ and must be written in terms of   $q$-Pochhammers, $q$-binomials, and $(q,A)$- dependent powers so that 
\begin{equation*}
    P^{\mathcal{K}}_r(A,q)=\sum_{k=0}^{r}{r \brack k}q^{-rk}(A q^r;q)_k\Tilde{c}^{\mathcal{K}}_k
\end{equation*} 
can be transformed into the following form to deduce  the corresponding quiver $Q_{\mathcal{K}}$:
\begin{eqnarray}
P^{\mathcal{K}}_r(A,q)&=&  
\sum_{\mathbf d}(-1)^{\sum_i\gamma_i d_i}\frac{q^{\sum_{i,j} C^{\mathcal{K}}_{i,j} d_i d_j} (q^2;q^2)_r}{\prod_{i=1}^m (q^2;q^2)_{d_i}}  \nonumber\\
&&q^{\sum
\alpha_i d_i} A^{\sum \beta_i d_i}.   \label{Pr-quiver1}
\end{eqnarray}
Here $C^{\mathcal{K}}_{i,j}$ is the quiver matrix and the variables $\alpha_i,~\beta_i$ and $\gamma_i$ are integer parameters. The set $\{d_i\}\equiv \mathbf d$ must obey $r = d_1 +d_2 \ldots + d_n$ with $d_i\geq 0$.
Even though such a transformation is motivated by comparing Ooguri-Vafa partition function\cite{Ooguri:2000tr} with the motivic generating series\cite{MR2889742,MR2851153,MR2956038}, it is still a hard problem to obtain $\Tilde{c}^{\mathcal{K}}_k$ for any knot.
\end{enumerate}
Note that the quadratic power of $q$ depends on $C^{\mathcal{K}}_{i,j}$ and it is independent of  $A=q^N$.
Hence, we will work with the colored Jones polynomials $J_r(\mathcal{K},q)$ of a knot $\mathcal{K}$ to extract its quiver matrix using the reverse engineering techniques of MMR formalism replacing $A\rightarrow q^2$ in eqn.(\ref{Pr-quiver1})\footnote{ Theorem 1.1, in  Ref.\cite{Stosic:2017wno} indicates the colored Jones polynomials of rational links also admit generating functions in quiver form.}.  

The plan of the paper is as follows: In section \ref{sec2}, we briefly discuss the colored Jones polynomials of double twist knot $K(p,-m)$ obtained from the reverse engineering techniques of MMR expansion. In section \ref{sec3}, we conjecture $C^{\mathcal{K}}_{i,j}$ for $\mathcal K = K(p,-m)$ and validate it for some double twist knots.  We conclude in section \ref{sec4}
summarising our results, enumerating some  open questions and
future directions.


\section{ Double twist knots}    \label{sec2}
We have listed some of the double twist knots $K(p,-m)$ in Table \ref{table:1}. 
\begin{table}[htp]
\begin{center}
\begin{tabular}{|c|c|}
\hline
$K(p,-m)$& Knots \\
\hline
$K(p,-1)$ & Twist Knots \\
\hline
K(2,-2)& $8_3$\\
\hline
K(3,-2)& $10_3$ \\
\hline
\end{tabular}
\end{center}
\caption{Examples of  double twist knots $K(p,-m)$} \label{table:1}
\end{table}

\begin{figure}[htbp]
\begin{center}
\includegraphics[scale=.3]{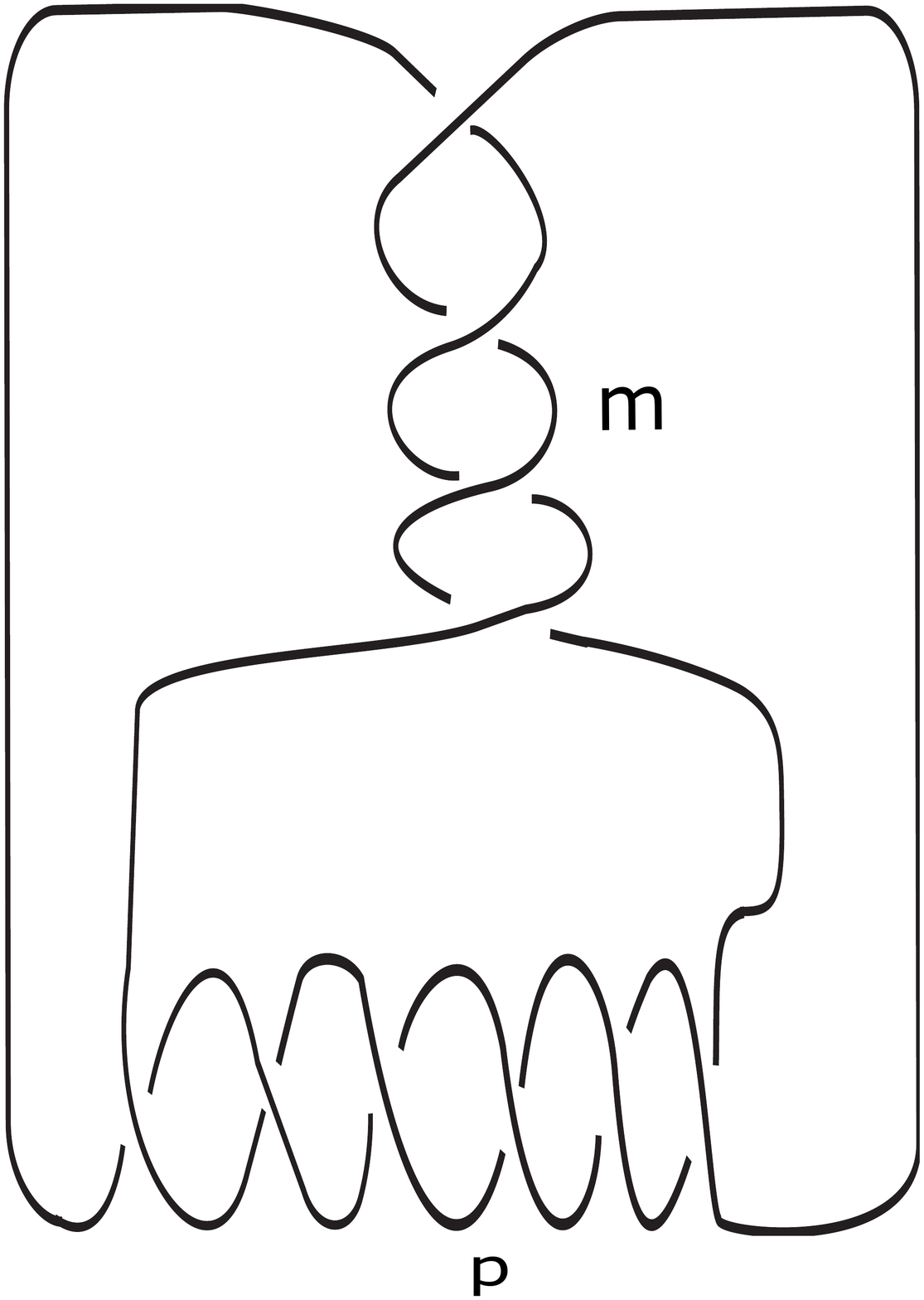}
\caption{Double twist knots $\newline$ ($m, p$ denote number of  full-twists)}\label{DT}
\end{center}
\end{figure}
As these double twist knots belong to arborescent family, the symmetric $r$-colored HOMFLY-PT polynomials can be obtained for every $r$ from Chern-Simons theory\cite{Devi:1993ue,Nawata:2012vk}. In fact, colored HOMFLY-PT for arbitrary $r$ in closed form is given in Ref. \cite{chen2021cyclotomic}. Hence, our aim is not to reconstruct  $r$-colored HOMFLY-PT for double twist knots. We will now present the reverse engineering of MMR formalism (\ref{mmr-intro}) to rewrite $r$-colored Jones as a motivic series to extract the matrix of the quiver $Q_{K(p,-m)}$. 
\subsection{Colored HOMFLY-PT polynomials for a class of Double twist knots $K(p,-m)$ }\label{sec2}
For given positive integers $p$ and $m$, the Alexander polynomial of a double twist knot of type $K(p,-m)$ takes the form
\begin{equation}
    \Delta^{K(p,-m)}(x) = 1 - (p ~m) X .    \label{Alexander-twist-I}
\end{equation}
Here $X=\frac{(1-x)^2}{x}$. Such a linear expression  appeared in many knots \cite{Banerjee_2020}, suggesting the
inverse binomial expansion to take the following form:
\begin{eqnarray}
    \frac{1}{\Delta^{K(p,-m)}(x)^{N-1}}& =& \sum_{0 \le k_1 \le \ldots \le k_{2 m p} }^\infty  {N+k_{2 m p}-2 \choose k_{2 m p}}\nonumber\\
&&\prod_{i=1}^{2mp-1}{k_{i+1} \choose k_i}  \frac{(1-x)^{2k_{2 m p}}}{x^{k_{2 m p}}}.   \label{CBE}
\end{eqnarray}
Further, using the quantum deformation procedure discussed in \cite{Banerjee_2020} and taking $A\rightarrow q^2 ( N=2)$ in eqn.(\ref{CBE}), the colored Jones polynomial  can be written as 
\begin{eqnarray}
J_r(K(p,-m),q)&=&  
\sum_{d_1+d_2 +\ldots +d_{4mp+1}=r}(-1)^{\sum_i \gamma_i d_i} q^{\sum
\xi_i d_i}\nonumber\\ &&\frac{q^{\sum_{i,j} C^{K(p,-m)}_{i,j} d_i d_j} (q^2;q^2)_r}{\prod_{i=1}^m (q^2;q^2)_{d_i}},  \label{Pr-quiver}
\end{eqnarray}
where $C^{K(p,-m)}_{i,j}$ is $(4 pm+1) \times (4 pm+1)$  matrix for quiver $Q_{K(p,-m)}$. It is worth noting that $\xi_i = \alpha_i + 2 \beta_i$ and $\gamma_i$ are integer parameters that can be determined by comparing them with  $r=1,2,3$ \cite{chen2021cyclotomic,lovejoy2017colored}.  By this approach, we explicitly  determined $\{\xi_i\},$
$\{\gamma_i\}$ parameters(\ref{Pr-quiver}) for $K_{(2,-2)}={\bf{8_3}}$ knot:
\begin{eqnarray} \label{83qui}
J_r(8_3,q)&=& \sum_{d_1+d_2 +\ldots +d_{17}=r}(-1)^{\sum_i \gamma_i d_i}\nonumber\\
&&(q^2;q^2)_r \frac{q^{\sum_{i,j} C_{i,j} d_i d_j }}{\prod_{i=1}^{17} (q^2;q^2)_{d_i}} q^{\sum
\xi_i d_i}\nonumber,
\end{eqnarray}
where\\ $
\sum_i \gamma_i d_i = d_{11} + d_{13} + d_{14} + d_{16} + d_{3} + d_{5} + d_{6} + d_{8} ~; $\\ $ \sum_i\xi_i d_i = (d_{11} - 2 d_{12} - d_{13} + d_{14} + 2 d_{15} + 3 d_{16} + 4 d_{17} \\- 2 d_2 - d_3 - 4 d_4 - 3 d_5 - d_6 + d_8 + 2d_9)$.\\ The quiver  matrix $C^{K(2,-2)}=C^{8_3}$ is as follows:
\vskip.1cm
\noindent
\resizebox{\linewidth}{!}{$\left(
\begin{array}{c|cccc|cccc|cccc|cccc}
 0 & -1 & -1 & -1 & -1 & 0 & 0 & 0 & 0 & -1 & -1 & -1 & -1 & 0 & 0 & 0 & 0 \\
 \hline
 -1 & -2 & -2 & -3 & -3 & -2 & -2 & -1 & -1 & -2 & -2 & -3 & -3 & -2 & -2 & -1 & -1 \\
 -1 & -2 & -1 & -2 & -2 & -1 & -1 & 0 & 0 & -1 & -1 & -2 & -2 & -1 & -1 & 0 & 0 \\
 -1 & -3 & -2 & -4 & -4 & -3 & -3 & -1 & -1 & -2 & -2 & -4 & -4 & -3 & -3 & -1 & -1 \\
 -1 & -3 & -2 & -4 & -3 & -2 & -2 & 0 & 0 & -1 & -1 & -3 & -3 & -2 & -2 & 0 & 0 \\
 \hline
 0 & -2 & -1 & -3 & -2 & -1 & -1 & 0 & 0 & -1 & -1 & -2 & -2 & -1 & -1 & 0 & 0 \\
 0 & -2 & -1 & -3 & -2 & -1 & 0 & 1 & 1 & 0 & 0 & -1 & -1 & 0 & 0 & 1 & 1 \\
 0 & -1 & 0 & -1 & 0 & 0 & 1 & 1 & 1 & 0 & 0 & 0 & 0 & 1 & 1 & 1 & 1 \\
 0 & -1 & 0 & -1 & 0 & 0 & 1 & 1 & 2 & 1 & 1 & 1 & 1 & 2 & 2 & 2 & 2 \\
 \hline
 -1 & -2 & -1 & -2 & -1 & -1 & 0 & 0 & 1 & 0 & 0 & -1 & -1 & 0 & 0 & 1 & 1 \\
 -1 & -2 & -1 & -2 & -1 & -1 & 0 & 0 & 1 & 0 & 1 & 0 & 0 & 1 & 1 & 2 & 2 \\
 -1 & -3 & -2 & -4 & -3 & -2 & -1 & 0 & 1 & -1 & 0 & -2 & -2 & -1 & -1 & 1 & 1 \\
 -1 & -3 & -2 & -4 & -3 & -2 & -1 & 0 & 1 & -1 & 0 & -2 & -1 & 0 & 0 & 2 & 2 \\
 \hline
 0 & -2 & -1 & -3 & -2 & -1 & 0 & 1 & 2 & 0 & 1 & -1 & 0 & 1 & 1 & 2 & 2 \\
 0 & -2 & -1 & -3 & -2 & -1 & 0 & 1 & 2 & 0 & 1 & -1 & 0 & 1 & 2 & 3 & 3 \\
 0 & -1 & 0 & -1 & 0 & 0 & 1 & 1 & 2 & 1 & 2 & 1 & 2 & 2 & 3 & 3 & 3 \\
 0 & -1 & 0 & -1 & 0 & 0 & 1 & 1 & 2 & 1 & 2 & 1 & 2 & 2 & 3 & 3 & 4 \\
  \end{array}
\right)$}
\vskip.1cm

\noindent
To give clarity to the readers, we present a step-by-step procedure for determining the quiver matrix for 
the knot $8_3$ in the Appendix. 
The polynomial invariants matches with the  closed form \cite{chen2021cyclotomic} for large value of $r$ as well confirming that the above $8_3$ quiver data is indeed correct. Such an  exercise for $K(2,-2)$  suggested that we could propose and conjecture $C^{K(p,-m)}$ for the double twist knot family. We discuss them in the following section.

\section{Knot-Quiver Correspondence of double twist knots $K(p,-m)$}   \label{sec3}
 We observe that the quiver matrix has a block structure by performing a similar analysis of the previous section for other examples of the double twist knots  $K(p, -m)$.  Our explicit computation suggests the following proposition.\\
\textbf{Proposition: }\\ \emph{The $r$-colored Jones polynomial for double twist knots $K(p,-m)$, with $p\geq m$, can be expressed in the quiver representation:
\begin{eqnarray}{\label{KQ}}
 J_{r}(K(p,-m);q)&=&\sum_{d_{1}+d_{2}+\ldots +d_{4pm+1}=r}(-1)^{\Lambda_{(p,-m)}}q^{\Xi_{(p,-m)}}\nonumber\\&&\frac{(q^2;q^2)_{r}}{\prod_{i=1}^{4pm+1}(q^2;q^2)_{d_i}}  q^{\sum_{i,j} C^{K(p,-m)}d_{i}d_j}\nonumber\\
\end{eqnarray}
where the linear term $\Xi_{(p,-m)}\equiv \sum_i \xi_i d_i$, phase factor $\Lambda_{(p,-m)}\equiv \sum_i \gamma_i d_i$.}\\
The block structure of the matrix $C^{K(p,-m)}$ for some examples lead to the following conjecture:\\
{\bf Conjecture:} \emph{ The generic structure of  the quiver  matrix will take the form}
\begin{equation}
\label{QUIVERDT}
C^{K(p,-m)}=\left(\arraycolsep=1.4pt\def\arraystretch{1.5}\begin{array}{c|c|c|c|c|c|c|c|c|c}
F_{0} & F_{k} & \Tilde{F}_{k} &\cdots & \cdots &F_{k} &\Tilde{F}_{k} &\cdots &F_{k} & \Tilde{F}_{k}\\
\hline
F_{k}^{\top} & U_{1} & R_1 &\Tilde{R}_1  & \cdots &\Tilde{R}_1&R_1  &\cdots&\Tilde{R}_1 & {R}_1  \\
\hline
\Tilde{F}_{k}^{\top} & R_{1}^{\top} & \Tilde{U}_{1} &T_1&\Tilde{T}_1  &\vdots &\vdots & \cdots &T_1 &\Tilde{T}_1 \\
\hline
{F}_{k}^{\top} & \Tilde{R}_{1}^{\top} & T_{1}^{\top} &U_2&\Tilde{R}_2  & \vdots&\vdots & \cdots &\Tilde{R}_2 &{R}_2 \\
\hline
\vdots & \vdots & \vdots & \vdots & \ddots &\vdots &\vdots &\cdots & \vdots & \vdots\\
\hline
F_{k}^{\top} & R_{1}^{\top} & \cdots & \cdots & \vdots &U_{i} &\cdots &\vdots & \Tilde{R}_{i} & {R}_{i}\\
\hline
\Tilde{F}_{k}^{\top} & \Tilde{R}_{1}^{\top} &  \cdots &R_{i}^{\top}  &  \cdots&&\Tilde{U}_{i} & \cdots& T_{i} & \tilde{T}_{i}\\
\hline
\vdots & \dots & \vdots & \vdots & \ddots &\vdots &\vdots &\ddots & \vdots & \vdots\\
\hline
F_{k}^{\top} &\Tilde{R}_{1}^{\top} & T_{1}^{\top} & \Tilde{R}_{2}^{\top}&\cdots&\vdots & \vdots& \cdots & U_{p} & R_{p}\\
\hline
\vspace{.2cm}
\Tilde{F}_{k}^{\top} & {R}_{1}^{\top} & \Tilde{T}_{1}^{\top} & {R}_{2}^{\top} & \cdots& \cdots&\vdots & \cdots & R_{p}^{\top} & \Tilde{U}_{p}
\end{array}\right),  
\end{equation}
where $X^{\top}$ stands for transposition of matrix $X$, the row matrices $$F_{k}=\left(-1,-1,\ldots -1,-1\right), \Tilde{F}_k=\left(0,0,\ldots 0,0\right),$$ of size $1\times 2m$ and $F_0=0$. 
Let $X_{k}$  denote the following set of  $2m \times 2m$ matrices :
\begin{equation}{\label{G}}
X_{k}=\{U_k,\Tilde{U}_{k},R_k,\Tilde{R}_{k},T_{k},\Tilde{T}_{k}\},
\end{equation}
where $k=1,2\ldots p$. All these matrices can be recursively obtained using
$$X_{k}=X_{k-1}+2 (k-1) J~~\forall k\geq 2,$$ where $J$ is a matrix of size $2m \times 2m$ where all the elements  are one. So, knowing the set $X_1$ is sufficient to determine the full quiver matrix $C^{K(p,-m)}$. 

It appears that the set $X_1$ for the simplest twist knot $K(p=1,-m=-1)\equiv 4_1$ will suffice to obtain $X_1$ for double twist knots $K(p,-m)$ as $K(p,-m) = K^*(m,-p)$ ($K^*$ denotes mirror image of the knot $K$).  However, our matrix conjecture 
assumes $p \geq m$.  Hence, our explicit computations of set $X_1$ for $m=2,3$ is not derivable  from the $C^{K(p,-1)}$. 

In the following subsections, we will give some examples to validate our proposition and conjecture. 
Specifically, we work out the $X_1$ set matrices for double twist knots $K(m,-m)$ for $m =1,2,3$. This set is sufficient to 
obtain the explicit quiver presentations for all the double twist knots $K(p, -m)$ where $m =1,2,3$. 
\subsection{Knot-Quiver correspondence for  $K(p,-1)$}
$K(p,-1)$ are known in the literature as `twist knots' which is the simplest class of double twist knots.
In this case, we fix the parameter $m=1$ and vary the other parameter $p$. 
The simplest example, we consider $p=1$, i.e. $K(1,-1)={\bf 4_1}$ knot. Using eqn.(\ref{KQ}), we obtained the quiver form of ${\bf 4_1}$ as
\begin{eqnarray*}
J_{r}(4_1;q) &=& \sum_{d_1+\ldots+d_5=r}(-1)^{d_3+d_4} q^{-2 d_2 - 2 d_1 d_2 - 2 d_2^2 - d_3 + 2 d_5^2}\\&&\frac{(q^2;q^2)_{r}}{(q^2;q^2)_{d_1}(q^2;q^2)_{d_2}(q^2;q^2)_{d_3}(q^2;q^2)_{d_4}(q^2;q^2)_{d_5}}\\&& q^{- 2 d_1 d_3 - 4 d_2 d_3 - d_3^2 + d_4 - 
 2 d_2 d_4 + d_4^2 + 2 d_5 - 2 d_2 d_5 + 2 d_4 d_5}. 
\end{eqnarray*}
Thus, the quiver matrix
\begin{eqnarray*} C^{4_1}=\left(
\begin{array}{c|cc|cc}
 0 & -1 & -1 & 0 & 0  \\
 \hline
 -1 & -2 & -2 & -1 & -1 \\
 -1& -2 & -1 & 0& 0  \\
 \hline
 0 & -1 & 0 & 1 & 1  \\
 0 & -1 & 0 & 1 & 2  \\
 \end{array}
\right).
\end{eqnarray*}
Similarly, we obtained other matrices for $p=2, 3$ i.e
\begin{eqnarray*} C^{6_1}=\left(
\begin{array}{c|cc|cc|cc|cc}
 0 & -1 & -1 & 0 & 0 & -1 & -1 & 0 & 0 \\
 \hline
 -1 & -2 & -2 & -1 & -1 & -2 & -2 & -1 & -1 \\
 -1 & -2 & -1 & 0 & 0 & -1 & -1 & 0 & 0 \\
 \hline
 0 & -1 & 0 & 1 & 1 & 0 & 0 & 1 & 1 \\
 0 & -1 & 0 & 1 & 2 & 1 & 1 & 2 & 2 \\
 \hline
 -1 & -2 & -1 & 0 & 1 & 0 & 0 & 1 & 1 \\
 -1 & -2 & -1 & 0 & 1 & 0 & 1 & 2 & 2 \\
 \hline
 0 & -1 & 0 & 1 & 2 & 1 & 2 & 3 & 3 \\
 0 & -1 & 0 & 1 & 2 & 1 & 2 & 3 & 4 \\
\end{array}
\right),
\end{eqnarray*}
\resizebox{\linewidth}{!}{$C^{8_1}=\left(
\begin{array}{c|cc|cc|cc|cc|cc|cc}
 0 & -1 & -1 & 0 & 0 & -1 & -1 & 0 & 0 & -1 & -1 & 0 & 0 \\
 \hline
 -1 & -2 & -2 & -1 & -1 & -2 & -2 & -1 & -1 & -2 & -2 & -1 & -1 \\
 -1 & -2 & -1 & 0 & 0 & -1 & -1 & 0 & 0 & -1 & -1 & 0 & 0 \\
 \hline
 0 & -1 & 0 & 1 & 1 & 0 & 0 & 1 & 1 & 0 & 0 & 1 & 1 \\
 0 & -1 & 0 & 1 & 2 & 1 & 1 & 2 & 2 & 1 & 1 & 2 & 2 \\
 \hline
 -1 & -2 & -1 & 0 & 1 & 0 & 0 & 1 & 1 & 0 & 0 & 1 & 1 \\
 -1 & -2 & -1 & 0 & 1 & 0 & 1 & 2 & 2 & 1 & 1 & 2 & 2 \\
 \hline
 0 & -1 & 0 & 1 & 2 & 1 & 2 & 3 & 3 & 2 & 2 & 3 & 3 \\
 0 & -1 & 0 & 1 & 2 & 1 & 2 & 3 & 4 & 3 & 3 & 4 & 4 \\
 \hline
 -1 & -2 & -1 & 0 & 1 & 0 & 1 & 2 & 3 & 2 & 2 & 3 & 3 \\
 -1 & -2 & -1 & 0 & 1 & 0 & 1 & 2 & 3 & 2 & 3 & 4 & 4 \\
 \hline
 0 & -1 & 0 & 1 & 2 & 1 & 2 & 3 & 4 & 3 & 4 & 5 & 5 \\
 0 & -1 & 0 & 1 & 2 & 1 & 2 & 3 & 4 & 3 & 4 & 5 & 6 \\
\end{array}
\right).$}\\
\vskip.1cm
\noindent
These three examples confirm our conjecture for $m=1$.
 For clarity, the explicit quiver matrix for any  twist knot $K(p,-1)$ is
\begin{equation}
C^{K(p,-1)}=\left(\arraycolsep=1.4pt\def\arraystretch{1.5}\begin{array}{c|c|c|c|c|c|c|c|c|c}
F_{0} & F_{k} & \Tilde{F}_{k} &\cdots & \cdots &F_{k} &\Tilde{F}_{k} &\cdots &F_{k} & \Tilde{F}_{k}\\
\hline
F_{k}^{\top} & U_{1} & R_1 &\Tilde{R}_1  & \cdots &\Tilde{R}_1&R_1  &\cdots&\Tilde{R}_1 & {R}_1  \\
\hline
\Tilde{F}_{k}^{\top} & R_{1}^{\top} & \Tilde{U}_{1} &T_1&\Tilde{T}_1  &\vdots &\vdots & \cdots &T_1 &\Tilde{T}_1 \\
\hline
{F}_{k}^{\top} & \Tilde{R}_{1}^{\top} & T_{1}^{\top} &U_2&\Tilde{R}_2 & \vdots&\vdots & \cdots &\Tilde{R}_2 &{R}_2 \\
\hline
\vdots & \vdots & \vdots & \vdots & \ddots &\vdots &\vdots &\cdots & \vdots & \vdots\\
\hline
F_{k}^{\top} & R_{1}^{\top} & \cdots & \cdots & \vdots &U_{i} &\cdots &\vdots & \Tilde{R}_{i} & {R}_{i}\\
\hline
\Tilde{F}_{k}^{\top} & \Tilde{R}_{1}^{\top} &  \cdots &R_{i}^{\top}  &  \cdots&\vdots&\Tilde{U}_{i} &\cdots & T_{i} & \tilde{T}_{i}\\
\hline
\vdots & \vdots & \vdots & \vdots & \ddots &\vdots &\vdots &\ddots & \vdots & \vdots\\
\hline
F_{k}^{\top} &\Tilde{R}_{1}^{\top} & T_{1}^{\top} & \Tilde{R}_{2}^{\top}&\cdots&\vdots & \vdots& \cdots & U_{p} & R_{p}\\
\hline
\Tilde{F}_{p}^{\top} & {R}_{1}^{\top} & \Tilde{T}_{1}^{\top} & {R}_{2}^{\top} & \cdots& \cdots&\vdots & \cdots & R_{p}^{\top} & \Tilde{U}_{p}
\end{array}\right),  
    \label{twist}
\end{equation}
where the generators $X_1\equiv (U_1, \tilde U_1,R_1, \tilde R_1, T_1, \tilde T_1)$ are as follows:
\begin{eqnarray*} U_{1}&=&\left(
\begin{array}{cc}
 -2 & -2   \\
 -2 & -1 \\
 \end{array}
\right),~\Tilde{U}_{1}=\left(
\begin{array}{cc}
 1&1  \\
 1& 2 \\
 \end{array}
\right),\\
 R_{1}&=&\left(
\begin{array}{cc}
 -1 & -1   \\
 0& 0 \\
 \end{array}
\right),~~\Tilde{R}_{1}=\left(
\begin{array}{cc}
 -2 & -2   \\
 -1  &  -1\\
 \end{array}
\right),\\
T_{1}&=&\left(
\begin{array}{cc}
 0 & 0   \\
 1 & 1 \\
 \end{array}
\right), ~~~~\Tilde{T}_{1}=\left(
\begin{array}{cc}
 1 &1   \\
 2 & 2\\
 \end{array}
\right),
\end{eqnarray*}
and 
$F_{k}=\begin{pmatrix}
    -1, & -1   
 \end{pmatrix}, \Tilde{F}_{k}=\begin{pmatrix}
    0, & 0   
 \end{pmatrix},~~\text{and} ~F_0=(0).$ 
Based on these calculations, we can infer the general expressions for the linear term $\Xi(p,-1)$ and the phase factor $\Lambda (p,-1)$ in proposition $(\ref{twist})$ for any given value of $p$:
\begin{eqnarray}{\label{LT1}}
\Xi_{(p,-1)}&=&2p d_{4p+1}+\sum_{i=1}^{p}\rho_{1}(i)d_{4 i-3}+\rho_{2}(i)d_{4 i-2}+\nonumber\\&&+\rho_{3}(i)d_{4 i-1}+\rho_{4}(i)d_{4 i},
\end{eqnarray} 
where~ $\rho_{1}(i)=2i-2, \rho_{2}(i)=2i-4, \rho_{3}(i)=-3+2i, \rho_{4}(i)=-1+2i$, and the phase factor is 
\begin{eqnarray}{\label{PH1}}
\Lambda_{(p,-1)}=\sum_{i=1}^{2p}d_{\frac{1}{2}((-1)^{i+1}+4i+1)}.
\end{eqnarray} 
These results agree with the quiver matrix of twist knots obtained in Ref.\cite{Kucharski:2017ogk}. 

\subsection{ Knot-Quiver correspondence for $K(p,-2)$}
We have already worked out $K(2,-2)\equiv 8_3$ knot in section \ref{sec2}. Further, we explicitly worked out colored 
Jones for $K(p=3,-2) \equiv 10_3$ and the quiver matrix elements  $C^{K(p=3,-2)}$ are presented in the Appendix.

Our matrix form for $p=3$ is consistent with our conjecture (\ref{QUIVERDT}), and the basic set of matrices  $X_1$ are:
\begin{eqnarray}{\label{GM2}} U_{1}&=&\left(
\begin{array}{cccc}
 -2 & -2 &-3 &-3    \\
 -2&-1 &-2&-2 \\
 -3 &-2  & -4 &-4\\
 -3 &-2 &-4&-3\\
 \end{array}
\right),\Tilde{U}_{1}=\left(
\begin{array}{cccc}
 -1 & -1 & 0&0 \\
 -1 & 0 & 1&1\\
 0 & 1 & 1&1 \\
 0 & 1 &1 &2 \\
 \end{array}
\right),\nonumber\\ R_{1}&=&\left(
\begin{array}{cccc}
 -2 & -2 & -1 & -1   \\
 -1 & -1 & 0&0 \\
 -3 & -3 & -1 &-1\\
 -2 & -2 & 0 &0\\
 \end{array}
\right),\Tilde{R}_{1}=\left(
\begin{array}{cccc}
 -2 & -2 & -3&-3 \\
 -1 & -1 & -2& -2\\
 -2 & -2 & -4&-4 \\
 -1 & -1 & -3&-3 \\
 \end{array}
\right),\nonumber\\ T_{1}&=&\left(
\begin{array}{cccc}
 -1 & -1 & -2 & -2   \\
 0 & 0 & -1&-1 \\
 0 & 0 & 0 &0\\
 1 &1  &1 &1\\
 \end{array}
\right),\Tilde{T}_{1}=\left(
\begin{array}{cccc}
 -1 & -1 & 0&0 \\
 0 &0 & 1& 1\\
 1 & 1 & 1&1 \\
 2 & 2 & 2&2 \\
 \end{array}
\right),
\end{eqnarray}
 \resizebox{\linewidth}{!}{$F_{k}=\begin{pmatrix}
    -1, & -1, & -1&-1   
 \end{pmatrix}$, $\Tilde{F}_{k}=\begin{pmatrix}
    0, & 0, & 0&0 
 \end{pmatrix}$, ~\\ \text{and}~$ F_0=(0)$}.
 We further worked out for $p=4,5$ as well and verified  our conjecture (\ref{QUIVERDT}) form obeyed. From these computations, we can deduce the general form of  the linear term $\Xi_{(p,-2)}$ and phase factor $\Lambda_{(p,-2)}$ in the proposition(\ref{KQ})  for arbitrary $p$ as:
\begin{eqnarray}{\label{LT2}}
\Xi_{(p,-2)}&=&\sum_{i}^{2p}(\tau_{1}(i)d_{4 i+1}+\tau_{2}(i)d_{4 i}+\tau_{3}(i)d_{4 i-1}\nonumber \\&&+\tau_{4}(i)d_{4 i-2}),
\end{eqnarray} 
where~ $\tau_{1}(i)=-2+2(-1)^i+i, ~\tau_{2}(i)=-3+2(-1)^i+i,~\tau_{3}(i)=-2+i, ~\tau_{4}(i)=-3+i$, and the phase factor is 
\be{\label{PH2}}
\Lambda_{(p,-2)}=\sum_{i=2}^{4p+1}d_{(-3 + 4 i - (-1)^{(\lfloor{\frac{i}{2}\rfloor})}\frac{1}{2})}.
\ee
Using the above data, we can write the colored Jones polynomial for any $K(p,-2)$ in quiver presentation  with the quiver matrix consistent with the conjecture (\ref{QUIVERDT}). So far, we have obtained the set of matrices $X_1$ for $m=1, 2$. With the hope of deducing gthe pattern for the set $X_1$ for any $m$, we will investigate double twist knots with $m=3$ in the following subsection.
\subsection{ Knot-Quiver correspondence for $K(p,-3)$}
Following reverse MMR, we could write the quiver presentation for $K(3,-3)$ and obtain the  quiver matrix $C^{K(3,-3)}$. The explicit matrix form is presented in the Appendix.
The generators ($X_1$) of the quiver matrix  can be read off comparing with the conjectured form (\ref{QUIVERDT}):\\
\resizebox{\linewidth}{!}{$ U_{1}=\left(
\begin{array}{cccccc}
 -2 & -2 & -3 &  -3 &-3&-3  \\
 -2& -1 & -2&-2&-2&-2\\
 -3 & -2 & -4&-4&-5&-5\\
-3 & -2 & -4 &-3&-4&-4\\
 -3 &-2 & -5 &-4&-6&-6\\
-3 &  -2& -5 &-4&-6&-5\\
 \end{array}
\right),
~~\Tilde{U}_{1}=\left(
\begin{array}{cccccc}
 -3 & -3 & -2&-2&0&0 \\
 -3 & -2 & -1& -1&1&1\\
 -2 & -1 & -1&-1&0&0 \\
 -2 & -1 &-1&0&1&1\\
 0 & 1 & 0&1 &1&1\\
 0 & 1 & 0&1&1&2\\
 \end{array}
\right)$,}\\
\resizebox{\linewidth}{!}{$R_{1}=\left(
\begin{array}{cccccc}
 -2 & -2 & -2&-2&-1&-1 \\
 -1 & -1 & -1&-1&0&0\\
 -4 & -4 & -3&-3&-1& -1\\
 -3 & -3 &-2&-2 &0&0\\
 -5& -5 & -3&-3 &-1&-1\\
 -4 & -4 &-2& -2&0&0\\
 \end{array}
\right),~~~
\Tilde{R}_{1}=\left(
\begin{array}{cccccc}
 -2 & -2 & -3&-3&-3&-3 \\
 -1 & -1 & -2& -2&-2&-2\\
 -2 & -2 & -4&-4&-5&-5 \\
 -1 & -1 & -3&-3 &-4&-4\\
 -2 & -2 & -4&-4 &-6&-6\\
 -1 & -1 & -3&-3&-5&-5\\
 \end{array}
\right),\nonumber
$}
\resizebox{\linewidth}{!}{$ T_{1}=\left(
\begin{array}{cccccc}
 -1 & -1 & -3&-3&-4&-4 \\
 0 & 0 & -2&-2&-3&-3\\
 -1 & -1 & -2&-2&-2&-2 \\
 0 & 0 & -1&-1 &-1&-1\\
 0 & 0 & 0&0 &0&0\\
 1 & 1 & 1&1 &1&1\\
 \end{array}
\right),~~
\Tilde{T}_{1}=\left(
\begin{array}{cccccc}
 -3 & -3 & -2&-2&0&0 \\
 -2 & -2 & -1& -1&1&1\\
 -1 & -1 & -1&-1&0&0 \\
 0 & 0 & 0&0 &1&1\\
 1 & 1 & 1&1 &1&1\\
 2 & 2 & 2&2 &2&2\\
 \end{array}
\right),$}\\

\noindent
\resizebox{\linewidth}{!}{$F_{k}=\begin{pmatrix}
    -1, & -1, & -1&-1&-1&-1   
 \end{pmatrix},~ \Tilde{F}_{k}=\begin{pmatrix}
    0, & 0, & 0&0  &0&0 
 \end{pmatrix} ,~\text{and}~ F_0=(0)$}. We have verified that our conjecture (\ref{QUIVERDT}) is true for $p=4, 5$. 
The linear term and phase factor in the proposition (\ref{KQ}) for $p=3,4$ are as follows:
\begin{eqnarray*}
\Xi_{(3,-3)}&=&-2d_2-d_3-4 d_4-3 d_5-6 d_6-5 d_7-3 d_8-\\&&2 d_9-d_{10}+d_{12}+2 d_{13}+d_{15}-2 d_{16}-\\&&d_{17}-4 d_{18}-
3 d_{19}-d_{20}+d_{22}+2 d_{23}+\\&&3 d_{24}+4 d_{25}+2 d_{26}+3 d_{27}+d_{29}-2 d_{30}\\&&-d_{31}+d_{32}+2 d_{33}+3 d_{34}+
4 d_{35}\\&&+5 d_{36}+6 d_{37}.\\
\Xi_{(4,-3)}&=&\Xi_{(3,-3)}+4 d_{38}+5 d_{39}+2 d_{40}+3 d_{41}+d_{43}+\\&&3 d_{44}+4 d_{45}+5 d_{46}+6 d_{47}+7 d_{48}+8 d_{49}.\\
\Lambda_{(3,-3)}&=&d_3+d_5+ d_7+d_8+d_{10}+ d_{12}+ d_{15}+ \\&&d_{17}+ d_{19}+d_{20}+ d_{22}+ d_{24}+ d_{27}\\&&+ d_{29}+ d_{31}+ d_{32} + d_{34}+d_{36}.\\
\Lambda_{(4,-3)}&=&\Lambda_{(3,-3)}+d_{39}+d_{41}+ d_{43}+ d_{44}+ d_{46}+d_{48}.
\end{eqnarray*}
Probably,there is a closed-form expression for $\Xi_{p,-3}$ and $\Lambda_{p,-3}$ for any $p$. We are not able
to infer the closed form from the above data. 

Ideally, it would be beneficial to find the set of matrices $X_1$ for any $m$ as well as the closed form for $\Xi_{p,-m}$ and $\Lambda_{p,-m}$. The size of the quiver matrix $(4mp+1) \times (4mp+1)$ makes the computations difficult.


\section{Conclusion and Discussion}\label{sec4}
Double twist knots $K(p,-m)$ depend on two full twist parameters $p,m \in \mathbb Z_+$ belong to the arborescent  family (see Fig.\ref{DT}).  Finding a quiver with matrix (\ref{QUIVERDT}) associated to each of the double twist knots was  attempted using reverse engineering of Melvin-Morton-Rozansky expansion. We observed the Alexander polynomial form to be $\Delta(X)=1-p m X$, almost similar to twist knots $K(p,-1)$ studied in Ref.\cite{Kucharski:2017ogk}. Comparing the structure of twist knot quiver, we put forth a proposition (\ref{KQ}) for colored Jones in a quiver presentation  as well as conjectured (\ref{QUIVERDT}) the structure of the quiver  matrix $C^{K(p,-m)}$ for any double twist knot $K(p,-m)$. 
We have explicitly worked out some double twist knots  to validate our proposition and the conjecture for $m=1,2,3$. Our detailed methodology shows the complexity of the equations to deduce a concise form for $\Xi_{p,-3}$ and $\Lambda_{p,-3}$.

We did attempt rewriting the superpolynomial for double twist knots $K(p,-m)$\cite{Wang:2020uz} in the quiver representation form using the $q$-multinomial identities\cite{Kucharski:2017ogk}. Unfortunately, we face computational difficulty even in determining the powers of $A$ dependence. That is., $\beta_1,\beta_2, \ldots$ in the motivic series(\ref{DT}). We are working on an  alternative method involving {\it homological diagrams} to obtain $r$-colored HOMFLY-PT in quiver representation\cite{DSVP:2023}. We hope to report these results soon\cite{DSVP:2023}.

It will be worth investigating whether our conjectured form for the
block structure in $C^{K(p,-m)}$ has any connections  to (i) the tangle operations and (ii) the holomorphic discs on the knot conormal. Probably, this could be another way to tackle quiver representation form for $r$-colored HOMFLY-PT.

Three-manifold invariants $F_{\mathcal K}(x=q^r,q)$ for knot complements $S^3\backslash \mathcal K$ can be deduced from the coefficients of the MMR expansions for $r$-colored Jones polynomial\cite{gukov2021two}. Subsequently, this invariant was refined  in Ref. \cite{Ekholm:2020lqy}. Similar to the knot-quiver correspondence, 
a motivic series for the  $F_{\mathcal K} (x,q)$ invariant was conjectured in Ref. \cite{Kucharski:2020}. Such a conjecture has been validated  for the torus knots of type (2,2p+1)\cite{Kucharski:2020}, the double twist knots of type $K(p,m), K(p+\frac{1}{2},-m)$\cite{Park:2020edg}. The large color behaviour of colored Jones polynomial was the starting point to tackle $F_{\mathcal K}(x,q)$ for positive braid knots in Ref.\cite{Park:2020edg}. Further, the refined version of these three-manifold invariant for the knot complements of positive braids is discussed in Ref.\cite{Wang_2021}.
We still face a  stumbling block in achieving a quiver form for  these double twist knots of type $K(p,m), K(p+\frac{1}{2},-m)$ and the motivic series $F_{K(p,-m)}(x,q)$. We probably need a radical approach to obtain such a motivic series form. We will report on these aspects and their refined version in future.

There is a pretzel family of knots whose $r$-colored Jones and HOMFLY-PT are known. It will be interesting exercise if we can explicitly write a quiver presentation and deduce the matrix for the pretzel. From our double twist knot results, it may be straightforward to attempt quiver presentation for knots whose   Alexander polynomial takes the form $\Delta(X)=1\pm (m_1 m_2 \ldots m_p )X~.$  We hope to address these problems in future. 

\vspace{0.5cm}

\begin{acknowledgments}
The work of VKS is supported by ``Tamkeen under the NYU Abu Dhabi Research Institute grant CG008 and  ASPIRE Abu Dhabi under Project AARE20-336''. VKS would like to thank P. Sulkowski, Q. Chen, Hisham Sati and Urs Schreiber for the helpful discussion. PR would like to thank SERB (MATRICS) MTR/2019/000956 funding, which enabled her to visit University of Warsaw and present these results. PR would also like to acknowledge the ICTP’s Associate programme, which helped her to complete this project during the visit as a senior associate. SC and PR would like to thank all the speakers and the organisers of the Learning workshop on BPS states and 3-manifolds for discussions and interactions on `knot-quiver' correspondence. BPM acknowledges the research grant for faculty under IoE Scheme (Number 6031) of Banaras Hindu University. AD would like to thank UGC for the research fellowship. P. R. would like to acknowledge the SPARC/2019-2020/
P2116/ project funding
\end{acknowledgments}

\appendix*
\section{Quiver matrix}
For clarity, we present the steps in obtaining the quiver representation for a knot $8_3$ in this Appendix.
The Alexander polynomial  in variable $X$ is $\Delta^{8_3}(X)=(1-4X)~.$
Such a form implied that we could perform the reverse  MMR method discussed\cite{Banerjee_2020} for $\Delta^K(X)= 1-pX$ 
with $p$ being integer. The $r$-colored Jones polynomials take the following interesting form:
\begin{eqnarray}   
J_{r}({\bf{8_3}};q) &=& \sum_{r\geq k_8\geq \ldots \geq k_1 \geq 0} (q^2;q^2)_{k_8} {r \brack  k_8} { k_8 \brack  k_7}{ k_7 \brack  k_6 }{k_6 \brack  k_5}\nonumber\\
~&~&{ k_5 \brack  k_4} { k_4 \brack  k_3}{ k_3 \brack  k_2 }{ k_2 \brack  k_1 } q^{\sum_{i,j=1}^p a_{ij} k_i k_j} 
q^{{\sum_{i=1}^p} b_i k_i}\nonumber\\
~&~&(-1)^{{\sum_{i=1}^p} c_i k_i}\nonumber
\end{eqnarray}
By comparing these polynomials with known $r$-colored Jones polynomials, we have successfully determined the unknown parameters ${a_{ij}, b_i, c_i}$.
The exact expression is
\begin{eqnarray*}
J_{r}({\bf{8_3}};q) &= & \sum_{r\geq k_8\geq\ldots \geq k_1 \geq 0} (-1)^{ k_2+k_4+ k_6}(q^2;q^2)_{k_8} {r \brack  k_8} \nonumber\\
&& \prod_{i=0}^{6}{k_{8-i}\brack k_{7-i}}q^{2 k_1+3 k_2+2 r k_2-2 k_1 k_2-2 k_3 k_8+2 k_5 k_8}\nonumber\\ && q^{-k_4-2 r k_4+2 k_1 k_4-2 k_3 k_4+k_4^2+2 k_5-2 k_1 k_5+2 k_3 k_5+3 k_6}\nonumber\\
&&q^{+2 r k_6-2 k_1 k_6+2 k_3 k_6-2 k_5 k_6+k_6^2-2 k_7+2 k_1 k_7-2 k_3 k_7}\nonumber\\
&&q^{+2 k_5 k_7-2 k_8-2 r k_8+2 k_1 k_8+k_2^2-2 k_3+2 k_1 k_3-2 k_7 k_8}.~\nonumber
\end{eqnarray*}
Using the $q$-binomial and $q$-Pochhammer identities discussed in Ref. \cite{Kucharski:2017ogk}, we could rewrite the $r$-colored Jones polynomial as
\begin{eqnarray}{\label{83jones1}}
J_{r}({\bf{8_3}};q)&=&\sum_{r, \mathbf k, \boldsymbol{\alpha}}\frac{(-1)^{k_2+k_4+k_6+\alpha_8}}{\prod_{i=0}^{6}(q^2;q^2)_{k_{8-i}-\alpha_{8-i}-k_{8-i-1}+\alpha_{8-i-1}}}\nonumber\\&&\frac{(q^2;q^2)_{r}}{(q^2;q^2)_{r-k_8}\prod_{i=0}^{6}(q^2;q^2)_{\alpha_{8-i}-\alpha_{8-i-1}}(q^2;q^2)_{k_1-\alpha_1}}\nonumber \\
&& \frac{1}{(q;q)_{\alpha_1}} q^{ \left(3k_2+2r k_2+k_2^2-k_4-2rk_4\right)+2\left(k_1-k_3+k_5-k_7\right)}\nonumber\\
&&q^{2\left(-k_1 k_2+k_1 k_3+k_1 k_4-k_3 k_4-k_1 k_5+k_3 k_5-k_1 k_6+k_3 k_6-k_5 k_6\right)}\nonumber\\
&& q^{\sum_{i=0}^{6}2(\alpha_{8-i} - \alpha_{7-i}) (k_{7-i} - \alpha_{7-i})+\alpha_8(\alpha_8+1)+k_4^2+3k_6 } \nonumber\\
&& \times  q^{\left(+2r k_6+k_6^2-2 k_8-k_8^2-2 r k_8+2(+k_1 k_7-k_3 k_7) \right)}\nonumber\\
&&q^{2(k_5 k_7+k_1 k_8-k_3 k_8+k_5 k_8-k_7
k_8)} ,\nonumber\\
\end{eqnarray}
where the summation variables in the first line must obey $r\geq k_8\geq \alpha_8 \geq k_7 \geq \alpha_7 \ldots \geq k_1\geq \alpha_1 \geq 0.$
By making the following substitutions \\
$\alpha_j = \sum_{i=0}^{j-1}d_{19-2j+2i}$ and $k_j = \sum_{i=0}^{2j-1}d_{18-2j+i}$\\ in eqn.(\ref{83jones1}), we obtain the motivic series quiver representation form:
\begin{eqnarray}
J_{r}({\bf{8_3}};q)&=& \sum_{d_1+d_2 \ldots+d_{17}=r}(-1)^{d_3+ d_5+ d_6+d_8+d_{11}+d_{13}+d_{14}+d_{16}}\nonumber\\
&& \frac{(q;q)_{r}q^{\sum C^{8_3}_{(i,j)} d_i d_j}}{\prod_{i=1}^{17}(q^2;q^2)_{d_{i}}}q^{\left(-2 d_2-d_3-4 d_4-3 d_5-d_6 \right)} \nonumber\\
&& q^{\left(d_8+2 d_9+d_{11}-2 d_{12}-d_{13}+d_{14}+2 d_{15} +3 d_{16}+4 d_{17} \right)}\nonumber.
\end{eqnarray}
We can read off the quiver  matrix elements $C^{8_3}$ from the above expression. See section\ref{sec2}, where we have presented the matrix.
The explicit matrix elements for $C^{K(3,-2)}$ and $C^{K(3,-3)}$ are presented below:\\
\resizebox{!}{4.2cm}{
$C^{(3,-2)}=\left(\begin{array}{c|cccc|cccc|cccc|cccc|cccc|cccc}
 0 & -1 & -1 & -1 & -1 & 0 & 0 & 0 & 0 & -1 & -1 & -1 & -1 & 0 & 0 & 0 & 0 & -1 & -1 & -1 & -1 & 0 & 0 & 0 & 0 \\
 \hline
 -1 & -2 & -2 & -3 & -3 & -2 & -2 & -1 & -1 & -2 & -2 & -3 & -3 & -2 & -2 & -1 & -1 & -2 & -2 & -3 & -3 & -2 & -2 & -1 & -1 \\
 -1 & -2 & -1 & -2 & -2 & -1 & -1 & 0 & 0 & -1 & -1 & -2 & -2 & -1 & -1 & 0 & 0 & -1 & -1 & -2 & -2 & -1 & -1 & 0 & 0 \\
 -1 & -3 & -2 & -4 & -4 & -3 & -3 & -1 & -1 & -2 & -2 & -4 & -4 & -3 & -3 & -1 & -1 & -2 & -2 & -4 & -4 & -3 & -3 & -1 & -1 \\
 -1 & -3 & -2 & -4 & -3 & -2 & -2 & 0 & 0 & -1 & -1 & -3 & -3 & -2 & -2 & 0 & 0 & -1 & -1 & -3 & -3 & -2 & -2 & 0 & 0 \\
 \hline
 0 & -2 & -1 & -3 & -2 & -1 & -1 & 0 & 0 & -1 & -1 & -2 & -2 & -1 & -1 & 0 & 0 & -1 & -1 & -2 & -2 & -1 & -1 & 0 & 0 \\
 0 & -2 & -1 & -3 & -2 & -1 & 0 & 1 & 1 & 0 & 0 & -1 & -1 & 0 & 0 & 1 & 1 & 0 & 0 & -1 & -1 & 0 & 0 & 1 & 1 \\
 0 & -1 & 0 & -1 & 0 & 0 & 1 & 1 & 1 & 0 & 0 & 0 & 0 & 1 & 1 & 1 & 1 & 0 & 0 & 0 & 0 & 1 & 1 & 1 & 1 \\
 0 & -1 & 0 & -1 & 0 & 0 & 1 & 1 & 2 & 1 & 1 & 1 & 1 & 2 & 2 & 2 & 2 & 1 & 1 & 1 & 1 & 2 & 2 & 2 & 2 \\
 \hline
 -1 & -2 & -1 & -2 & -1 & -1 & 0 & 0 & 1 & 0 & 0 & -1 & -1 & 0 & 0 & 1 & 1 & 0 & 0 & -1 & -1 & 0 & 0 & 1 & 1 \\
 -1 & -2 & -1 & -2 & -1 & -1 & 0 & 0 & 1 & 0 & 1 & 0 & 0 & 1 & 1 & 2 & 2 & 1 & 1 & 0 & 0 & 1 & 1 & 2 & 2 \\
 -1 & -3 & -2 & -4 & -3 & -2 & -1 & 0 & 1 & -1 & 0 & -2 & -2 & -1 & -1 & 1 & 1 & 0 & 0 & -2 & -2 & -1 & -1 & 1 & 1 \\
 -1 & -3 & -2 & -4 & -3 & -2 & -1 & 0 & 1 & -1 & 0 & -2 & -1 & 0 & 0 & 2 & 2 & 1 & 1 & -1 & -1 & 0 & 0 & 2 & 2 \\
 \hline
 0 & -2 & -1 & -3 & -2 & -1 & 0 & 1 & 2 & 0 & 1 & -1 & 0 & 1 & 1 & 2 & 2 & 1 & 1 & 0 & 0 & 1 & 1 & 2 & 2 \\
 0 & -2 & -1 & -3 & -2 & -1 & 0 & 1 & 2 & 0 & 1 & -1 & 0 & 1 & 2 & 3 & 3 & 2 & 2 & 1 & 1 & 2 & 2 & 3 & 3 \\
 0 & -1 & 0 & -1 & 0 & 0 & 1 & 1 & 2 & 1 & 2 & 1 & 2 & 2 & 3 & 3 & 3 & 2 & 2 & 2 & 2 & 3 & 3 & 3 & 3 \\
 0 & -1 & 0 & -1 & 0 & 0 & 1 & 1 & 2 & 1 & 2 & 1 & 2 & 2 & 3 & 3 & 4 & 3 & 3 & 3 & 3 & 4 & 4 & 4 & 4 \\
 \hline
 -1 & -2 & -1 & -2 & -1 & -1 & 0 & 0 & 1 & 0 & 1 & 0 & 1 & 1 & 2 & 2 & 3 & 2 & 2 & 1 & 1 & 2 & 2 & 3 & 3 \\
 -1 & -2 & -1 & -2 & -1 & -1 & 0 & 0 & 1 & 0 & 1 & 0 & 1 & 1 & 2 & 2 & 3 & 2 & 3 & 2 & 2 & 3 & 3 & 4 & 4 \\
 -1 & -3 & -2 & -4 & -3 & -2 & -1 & 0 & 1 & -1 & 0 & -2 & -1 & 0 & 1 & 2 & 3 & 1 & 2 & 0 & 0 & 1 & 1 & 3 & 3 \\
 -1 & -3 & -2 & -4 & -3 & -2 & -1 & 0 & 1 & -1 & 0 & -2 & -1 & 0 & 1 & 2 & 3 & 1 & 2 & 0 & 1 & 2 & 2 & 4 & 4 \\
 \hline
 0 & -2 & -1 & -3 & -2 & -1 & 0 & 1 & 2 & 0 & 1 & -1 & 0 & 1 & 2 & 3 & 4 & 2 & 3 & 1 & 2 & 3 & 3 & 4 & 4 \\
 0 & -2 & -1 & -3 & -2 & -1 & 0 & 1 & 2 & 0 & 1 & -1 & 0 & 1 & 2 & 3 & 4 & 2 & 3 & 1 & 2 & 3 & 4 & 5 & 5 \\
 0 & -1 & 0 & -1 & 0 & 0 & 1 & 1 & 2 & 1 & 2 & 1 & 2 & 2 & 3 & 3 & 4 & 3 & 4 & 3 & 4 & 4 & 5 & 5 & 5 \\
 0 & -1 & 0 & -1 & 0 & 0 & 1 & 1 & 2 & 1 & 2 & 1 & 2 & 2 & 3 & 3 & 4 & 3 & 4 & 3 & 4 & 4 & 5 & 5 & 6 \\
\end{array}\right)$,}

\vspace{.2cm}
\resizebox{!}{4.3cm}{$C^{(3,-3)} =\left(
\begin{array}{c|cccccc|cccccc|cccccc|cccccc|cccccc|cccccc}
 0 & -1 & -1 & -1 & -1 & -1 & -1 & 0 & 0 & 0 & 0 & 0 & 0 & -1 & -1 & -1 & -1 & -1 & -1 & 0 & 0 & 0 & 0 & 0 & 0 & -1 & -1 & -1 & -1 & -1 & -1 & 0
& 0 & 0 & 0 & 0 & 0 \\
\hline
 -1 & -2 & -2 & -3 & -3 & -3 & -3 & -2 & -2 & -2 & -2 & -1 & -1 & -2 & -2 & -3 & -3 & -3 & -3 & -2 & -2 & -2 & -2 & -1 & -1 & -2 & -2 & -3 & -3 &
-3 & -3 & -2 & -2 & -2 & -2 & -1 & -1 \\
 -1 & -2 & -1 & -2 & -2 & -2 & -2 & -1 & -1 & -1 & -1 & 0 & 0 & -1 & -1 & -2 & -2 & -2 & -2 & -1 & -1 & -1 & -1 & 0 & 0 & -1 & -1 & -2 & -2 & -2
& -2 & -1 & -1 & -1 & -1 & 0 & 0 \\
 -1 & -3 & -2 & -4 & -4 & -5 & -5 & -4 & -4 & -3 & -3 & -1 & -1 & -2 & -2 & -4 & -4 & -5 & -5 & -4 & -4 & -3 & -3 & -1 & -1 & -2 & -2 & -4 & -4 &
-5 & -5 & -4 & -4 & -3 & -3 & -1 & -1 \\
 -1 & -3 & -2 & -4 & -3 & -4 & -4 & -3 & -3 & -2 & -2 & 0 & 0 & -1 & -1 & -3 & -3 & -4 & -4 & -3 & -3 & -2 & -2 & 0 & 0 & -1 & -1 & -3 & -3 & -4
& -4 & -3 & -3 & -2 & -2 & 0 & 0 \\
 -1 & -3 & -2 & -5 & -4 & -6 & -6 & -5 & -5 & -3 & -3 & -1 & -1 & -2 & -2 & -4 & -4 & -6 & -6 & -5 & -5 & -3 & -3 & -1 & -1 & -2 & -2 & -4 & -4 &
-6 & -6 & -5 & -5 & -3 & -3 & -1 & -1 \\
 -1 & -3 & -2 & -5 & -4 & -6 & -5 & -4 & -4 & -2 & -2 & 0 & 0 & -1 & -1 & -3 & -3 & -5 & -5 & -4 & -4 & -2 & -2 & 0 & 0 & -1 & -1 & -3 & -3 & -5
& -5 & -4 & -4 & -2 & -2 & 0 & 0 \\
\hline
 0 & -2 & -1 & -4 & -3 & -5 & -4 & -3 & -3 & -2 & -2 & 0 & 0 & -1 & -1 & -3 & -3 & -4 & -4 & -3 & -3 & -2 & -2 & 0 & 0 & -1 & -1 & -3 & -3 & -4 &
-4 & -3 & -3 & -2 & -2 & 0 & 0 \\
 0 & -2 & -1 & -4 & -3 & -5 & -4 & -3 & -2 & -1 & -1 & 1 & 1 & 0 & 0 & -2 & -2 & -3 & -3 & -2 & -2 & -1 & -1 & 1 & 1 & 0 & 0 & -2 & -2 & -3 & -3
& -2 & -2 & -1 & -1 & 1 & 1 \\
 0 & -2 & -1 & -3 & -2 & -3 & -2 & -2 & -1 & -1 & -1 & 0 & 0 & -1 & -1 & -2 & -2 & -2 & -2 & -1 & -1 & -1 & -1 & 0 & 0 & -1 & -1 & -2 & -2 & -2 &
-2 & -1 & -1 & -1 & -1 & 0 & 0 \\
 0 & -2 & -1 & -3 & -2 & -3 & -2 & -2 & -1 & -1 & 0 & 1 & 1 & 0 & 0 & -1 & -1 & -1 & -1 & 0 & 0 & 0 & 0 & 1 & 1 & 0 & 0 & -1 & -1 & -1 & -1 & 0 &
0 & 0 & 0 & 1 & 1 \\
 0 & -1 & 0 & -1 & 0 & -1 & 0 & 0 & 1 & 0 & 1 & 1 & 1 & 0 & 0 & 0 & 0 & 0 & 0 & 1 & 1 & 1 & 1 & 1 & 1 & 0 & 0 & 0 & 0 & 0 & 0 & 1 & 1 & 1 & 1 & 1
& 1 \\
 0 & -1 & 0 & -1 & 0 & -1 & 0 & 0 & 1 & 0 & 1 & 1 & 2 & 1 & 1 & 1 & 1 & 1 & 1 & 2 & 2 & 2 & 2 & 2 & 2 & 1 & 1 & 1 & 1 & 1 & 1 & 2 & 2 & 2 & 2 & 2
& 2 \\
\hline
 -1 & -2 & -1 & -2 & -1 & -2 & -1 & -1 & 0 & -1 & 0 & 0 & 1 & 0 & 0 & -1 & -1 & -1 & -1 & 0 & 0 & 0 & 0 & 1 & 1 & 0 & 0 & -1 & -1 & -1 & -1 & 0 &
0 & 0 & 0 & 1 & 1 \\
 -1 & -2 & -1 & -2 & -1 & -2 & -1 & -1 & 0 & -1 & 0 & 0 & 1 & 0 & 1 & 0 & 0 & 0 & 0 & 1 & 1 & 1 & 1 & 2 & 2 & 1 & 1 & 0 & 0 & 0 & 0 & 1 & 1 & 1 &
1 & 2 & 2 \\
 -1 & -3 & -2 & -4 & -3 & -4 & -3 & -3 & -2 & -2 & -1 & 0 & 1 & -1 & 0 & -2 & -2 & -3 & -3 & -2 & -2 & -1 & -1 & 1 & 1 & 0 & 0 & -2 & -2 & -3 & -3
& -2 & -2 & -1 & -1 & 1 & 1 \\
 -1 & -3 & -2 & -4 & -3 & -4 & -3 & -3 & -2 & -2 & -1 & 0 & 1 & -1 & 0 & -2 & -1 & -2 & -2 & -1 & -1 & 0 & 0 & 2 & 2 & 1 & 1 & -1 & -1 & -2 & -2
& -1 & -1 & 0 & 0 & 2 & 2 \\
 -1 & -3 & -2 & -5 & -4 & -6 & -5 & -4 & -3 & -2 & -1 & 0 & 1 & -1 & 0 & -3 & -2 & -4 & -4 & -3 & -3 & -1 & -1 & 1 & 1 & 0 & 0 & -2 & -2 & -4 & -4
& -3 & -3 & -1 & -1 & 1 & 1 \\
 -1 & -3 & -2 & -5 & -4 & -6 & -5 & -4 & -3 & -2 & -1 & 0 & 1 & -1 & 0 & -3 & -2 & -4 & -3 & -2 & -2 & 0 & 0 & 2 & 2 & 1 & 1 & -1 & -1 & -3 & -3
& -2 & -2 & 0 & 0 & 2 & 2 \\
\hline
 0 & -2 & -1 & -4 & -3 & -5 & -4 & -3 & -2 & -1 & 0 & 1 & 2 & 0 & 1 & -2 & -1 & -3 & -2 & -1 & -1 & 0 & 0 & 2 & 2 & 1 & 1 & -1 & -1 & -2 & -2 & -1
& -1 & 0 & 0 & 2 & 2 \\
 0 & -2 & -1 & -4 & -3 & -5 & -4 & -3 & -2 & -1 & 0 & 1 & 2 & 0 & 1 & -2 & -1 & -3 & -2 & -1 & 0 & 1 & 1 & 3 & 3 & 2 & 2 & 0 & 0 & -1 & -1 & 0 &
0 & 1 & 1 & 3 & 3 \\
 0 & -2 & -1 & -3 & -2 & -3 & -2 & -2 & -1 & -1 & 0 & 1 & 2 & 0 & 1 & -1 & 0 & -1 & 0 & 0 & 1 & 1 & 1 & 2 & 2 & 1 & 1 & 0 & 0 & 0 & 0 & 1 & 1 & 1
& 1 & 2 & 2 \\
 0 & -2 & -1 & -3 & -2 & -3 & -2 & -2 & -1 & -1 & 0 & 1 & 2 & 0 & 1 & -1 & 0 & -1 & 0 & 0 & 1 & 1 & 2 & 3 & 3 & 2 & 2 & 1 & 1 & 1 & 1 & 2 & 2 & 2
& 2 & 3 & 3 \\
 0 & -1 & 0 & -1 & 0 & -1 & 0 & 0 & 1 & 0 & 1 & 1 & 2 & 1 & 2 & 1 & 2 & 1 & 2 & 2 & 3 & 2 & 3 & 3 & 3 & 2 & 2 & 2 & 2 & 2 & 2 & 3 & 3 & 3 & 3 & 3
& 3 \\
 0 & -1 & 0 & -1 & 0 & -1 & 0 & 0 & 1 & 0 & 1 & 1 & 2 & 1 & 2 & 1 & 2 & 1 & 2 & 2 & 3 & 2 & 3 & 3 & 4 & 3 & 3 & 3 & 3 & 3 & 3 & 4 & 4 & 4 & 4 & 4
& 4 \\
\hline
 -1 & -2 & -1 & -2 & -1 & -2 & -1 & -1 & 0 & -1 & 0 & 0 & 1 & 0 & 1 & 0 & 1 & 0 & 1 & 1 & 2 & 1 & 2 & 2 & 3 & 2 & 2 & 1 & 1 & 1 & 1 & 2 & 2 & 2 &
2 & 3 & 3 \\
 -1 & -2 & -1 & -2 & -1 & -2 & -1 & -1 & 0 & -1 & 0 & 0 & 1 & 0 & 1 & 0 & 1 & 0 & 1 & 1 & 2 & 1 & 2 & 2 & 3 & 2 & 3 & 2 & 2 & 2 & 2 & 3 & 3 & 3 &
3 & 4 & 4 \\
 -1 & -3 & -2 & -4 & -3 & -4 & -3 & -3 & -2 & -2 & -1 & 0 & 1 & -1 & 0 & -2 & -1 & -2 & -1 & -1 & 0 & 0 & 1 & 2 & 3 & 1 & 2 & 0 & 0 & -1 & -1 & 0
& 0 & 1 & 1 & 3 & 3 \\
 -1 & -3 & -2 & -4 & -3 & -4 & -3 & -3 & -2 & -2 & -1 & 0 & 1 & -1 & 0 & -2 & -1 & -2 & -1 & -1 & 0 & 0 & 1 & 2 & 3 & 1 & 2 & 0 & 1 & 0 & 0 & 1 &
1 & 2 & 2 & 4 & 4 \\
 -1 & -3 & -2 & -5 & -4 & -6 & -5 & -4 & -3 & -2 & -1 & 0 & 1 & -1 & 0 & -3 & -2 & -4 & -3 & -2 & -1 & 0 & 1 & 2 & 3 & 1 & 2 & -1 & 0 & -2 & -2 &
-1 & -1 & 1 & 1 & 3 & 3 \\
 -1 & -3 & -2 & -5 & -4 & -6 & -5 & -4 & -3 & -2 & -1 & 0 & 1 & -1 & 0 & -3 & -2 & -4 & -3 & -2 & -1 & 0 & 1 & 2 & 3 & 1 & 2 & -1 & 0 & -2 & -1 &
0 & 0 & 2 & 2 & 4 & 4 \\
\hline
 0 & -2 & -1 & -4 & -3 & -5 & -4 & -3 & -2 & -1 & 0 & 1 & 2 & 0 & 1 & -2 & -1 & -3 & -2 & -1 & 0 & 1 & 2 & 3 & 4 & 2 & 3 & 0 & 1 & -1 & 0 & 1 & 1
& 2 & 2 & 4 & 4 \\
 0 & -2 & -1 & -4 & -3 & -5 & -4 & -3 & -2 & -1 & 0 & 1 & 2 & 0 & 1 & -2 & -1 & -3 & -2 & -1 & 0 & 1 & 2 & 3 & 4 & 2 & 3 & 0 & 1 & -1 & 0 & 1 & 2
& 3 & 3 & 5 & 5 \\
 0 & -2 & -1 & -3 & -2 & -3 & -2 & -2 & -1 & -1 & 0 & 1 & 2 & 0 & 1 & -1 & 0 & -1 & 0 & 0 & 1 & 1 & 2 & 3 & 4 & 2 & 3 & 1 & 2 & 1 & 2 & 2 & 3 & 3
& 3 & 4 & 4 \\
 0 & -2 & -1 & -3 & -2 & -3 & -2 & -2 & -1 & -1 & 0 & 1 & 2 & 0 & 1 & -1 & 0 & -1 & 0 & 0 & 1 & 1 & 2 & 3 & 4 & 2 & 3 & 1 & 2 & 1 & 2 & 2 & 3 & 3
& 4 & 5 & 5 \\
 0 & -1 & 0 & -1 & 0 & -1 & 0 & 0 & 1 & 0 & 1 & 1 & 2 & 1 & 2 & 1 & 2 & 1 & 2 & 2 & 3 & 2 & 3 & 3 & 4 & 3 & 4 & 3 & 4 & 3 & 4 & 4 & 5 & 4 & 5 & 5
& 5 \\
 0 & -1 & 0 & -1 & 0 & -1 & 0 & 0 & 1 & 0 & 1 & 1 & 2 & 1 & 2 & 1 & 2 & 1 & 2 & 2 & 3 & 2 & 3 & 3 & 4 & 3 & 4 & 3 & 4 & 3 & 4 & 4 & 5 & 4 & 5 & 5
& 6 \\
\end{array}\right)$.}\\

\clearpage

\bibliographystyle{apsrev4-1}
\bibliography{ref.bib}

\end{document}